\documentstyle[12pt,a4,epsf]{article}
\newcommand{\be}{\begin{equation}}
\newcommand{\ee}{\end{equation}}
\newcommand{\etal}{{\it et al.}}
\newcommand{\hmp}{h^{-1}Mpc}
\newcommand{\bef}{\begin{figure}}
\newcommand{\eef}{\end{figure}}
 
\def\spose#1{\hbox to 0pt{#1\hss}}
\def\ltapprox{\mathrel{\spose{\lower 3pt\hbox{$\mathchar"218$}}
 \raise 2.0pt\hbox{$\mathchar"13C$}}}
\def\gtapprox{\mathrel{\spose{\lower 3pt\hbox{$\mathchar"218$}}
 \raise 2.0pt\hbox{$\mathchar"13E$}}}
\def\inapprox{\mathrel{\spose{\lower 3pt\hbox{$\mathchar"218$}}
 \raise 2.0pt\hbox{$\mathchar"232$}}}

\begin{document}
\centerline{}
\centerline{\LARGE Comment on the paper by L. Guzzo} 
\centerline{}
\centerline{\LARGE "Is the Universe homogeneous ?"}
\centerline{}

\centerline{}
\centerline{\Large 9 January 1998}
\centerline{}
\centerline{}
\centerline{ \Large F. Sylos Labini$^{1,2,3}$, M. Montuori$^{1}$ }
\centerline{}
\centerline{ \Large and  L. Pietronero$^{1}$}
\centerline{}
\centerline{\footnotesize $^{1}$ Dipartimento di Fisica, Universit\`a di Roma
``La Sapienza''}
\centerline{\footnotesize P.le A. Moro 2, I-00185 Roma, Italy}
\centerline{\footnotesize and INFM, Sezione di Roma 1}
\centerline{}
\centerline{\footnotesize  $^{2}$  D\'ept.~de Physique Th\'eorique,
 Universit\'e de Gen\`eve} 
\centerline{\footnotesize 24, Quai E. Ansermet, CH-1211 Gen\`eve, Switzerland}
\centerline{}
\centerline{\footnotesize  $^{3}$  Observatoire de Paris DEMIRM} 
\centerline{\footnotesize 61, Rue de l'Observatoire 75014 Paris, France}

\begin{abstract}
 This comment is in response to the paper by L. Guzzo recently
 appeared in ``New Astronomy'' related to our work.
 The subject of the discussion concerns 
 the correlation properties of galaxy distribution in the 
 available 3-d samples. There is a general agreement that galaxy 
 structures
 exhibit fractal properties, at least up to some scale.
 However the presence of an eventual
 crossover towards homogenization, as well as the exact value of
 the fractal dimension, are still matter of debate. 
 Here we briefly summarize our point of view by 
 discussing three main topics. The first one is methodological,
 i.e. we clarify 
 which are the correct methods to detect the 
 real correlations properties of the 3-d galaxy distribution. Then 
 we discuss the results of the analysis of several samples
 in two ranges of scales. In the first range of scale, 
 below $100 \div 200 \hmp$,
 the statistical quality of the data is rather good, and we find that 
 galaxy distribution has fractal properties with $D \approx 2$. 
 At larger distances the statistical robustness of the present
 data is weaker, but, we show that there is evidence 
 for a continuation of the fractal  behavior, without 
 any tendency towards homogenization.

\end{abstract}

\newpage

\section{Introduction}

 In a recent paper L. Guzzo \cite{guzzo}
 has exposed his arguments in favor of the 
 homogeneity of 
 galaxy distribution in the available three dimensional samples.
 This paper takes its origin from a discussion between 
 Dr. Guzzo and one of us (F.S.L.), held  during 
 the fifth Italian National Cosmology Meeting (Dec. 1996).
 This debate followed the one occurred in the Conference
 "Critical Dialogues in Cosmology" between Prof. M. Davis and 
 L.P. (June 1996) (see \cite{pmsl97} and \cite{dav97}).
 Moreover as a supplement to the debate 
 between L.P. and M. Davis, Prof. Peebles has sent
 a circular letter where he exposes his arguments 
 in favor of homogeneity, followed by a similar 
 letter that L.P. sent to Prof. Peebles. 
 All this material is now available at the {\it home page:
 http://www.phys.uniroma1.it/DOCS/PIL/pil.html}.

 In this comment we briefly summarize,
 in a colloquial way, our opinion
 about the arguments of Dr. Guzzo 
 and we refer the reader to a recent  review \cite{slmp97}
 for a more comprehensive and detailed discussion of the subject.

 Dr. Guzzo puts a special emphasis on the fact that 
 the identification of the scale at which galaxy distribution
 becomes homogeneous is one of the major topics of modern Cosmology.
 Homogeneity is, in fact, 
 the basic assumption of the current theory of galaxy formation,
 and in general is the cornerstone of any cosmological theory.
 Moreover it has been elevated to the role of a principle by the
 Cosmological Principle (e.g. \cite{pee93}).
 We have discussed in several papers
 \cite{cp92} \cite{sl94} \cite{slmp97}
 that homogeneity is a very strong a-priori assumption, although it has
 been a quite reasonable one  
 up to the compilation of large redshift surveys.
 As Mandelbrot stressed several times \cite{man82} \cite{man97}, 
 the assumption of local isotropy, without analyticity of matter  
 distribution, ensures as well the equivalence of all the occupied points 
 in the universe.
 Hence from a conceptual point of view, an isotropic 
 fractal distribution 
 is completely compatible with the requirement of having no  
 special directions or positions in the universe.
 
 Therefore, it is a crucial challenge in nowadays Cosmology
 to test whether the matter distribution is analytical or not, or 
 in other words, whether there exists a characteristic length in galaxy
 (and cluster) distributions. To this end a particular care must be
 used in the discussion of the correlation properties of galaxy samples.
 As Guzzo does, we focus here on the properties of redshift 
 samples, and we refer the reader to \cite{slmp97}  \cite{msl97} and
 for a more detailed discussions of the angular properties of
 real distributions, and to \cite{man82}  \cite{durrer}
 \cite{pfen} \cite{man97} for those of artificial fractals.

  Dr. Guzzo correctly pointed out that 
 {\it "the specific scale at which the galaxy 
 distribution apparently
 turns to homogeneity is dangerously close to the 
 size of the largest sample presently available"},  
 i.e. $\sim 100 \div 200 \hmp$. However he claims that 
 at larger distances there are enough evidences to conclude
 that homogeneity is well established.
 Here we will argue that there are well defined 
 and ample evidences that the transition scale to 
 homogeneity is larger than $100 \div 200 \hmp$, 
 and that up to this distance galaxy (and cluster) distribution shows
 well defined power law correlations, corresponding to a
 fractal dimension $ D \approx 2$. 
 Hence we partially agree with Dr. Guzzo
 about the small scale properties of galaxy distribution.
 However, we find in his arguments a lack of
 discussion of the implications of this result,
 especially for what concerns the luminosity segregation effect
 and, more in general, for the methods used to characterize
 galaxy correlations at these distances. 
 For example if galaxy distribution becomes homogeneous at
 $\sim 100 \hmp$, what is the meaning of the ``correlation lengths''
 of galaxies and clusters ? 
 These are important
 points having a number of implication, which we consider
 here in more detail.

 Moreover we discuss our point of view, about galaxy and cluster
 distribution at scales larger than $100 \div 200 \hmp$, 
 trying to clarify the statistical robustness of our results.
 In this range of length scales we disagree with Dr. Guzzo.
 In particular we show that,
 although the statistical quality of the data at these distances is
 weaker than at smaller scales, there is no any evidence for
 homogenization in any of the 3-d samples published up to now, and
 on the contrary there are evidence which support the continuation of
 the fractal behavior with dimension $D \approx 2$ found at smaller
 distances.

\section {A methodological point}

 There is a general agreement that galaxy distribution 
 exhibits fractal behavior up to a certain scale 
 (e.g.\cite{man82} \cite{pee80} \cite{cp92} \cite{pee93}).
 The eventual presence of a transition scale towards homogenization 
 and the exact value of the fractal dimension are matters
 of the present debate \cite{pmsl97} \cite{dav97}. Given this situation,
 it is  first of all crucial to establish which are 
 the  statistical methods suitable and appropriate to characterize 
 the correlation properties of galaxy distribution. 
 In particular we answer to the following
 question: which are the statistical 
 tools able to eventually identify 
 the homogeneity scale and to measure the 
 correlation exponent in the correct way ?

 The proper methods to characterize irregular as 
 well as regular distributions have been correctly illustrated
 by Dr. Guzzo in Sec.3, and we refer to \cite{cp92} and
 \cite{slmp97} for a more detailed and exhaustive
 discussion. The basic point is that, as far as a system 
 shows power law correlations, the usual $\xi(r)$ analysis 
 (e.g. \cite{pee80}) gives an incorrect result, since it is 
 based on the a-priori assumption of homogeneity. 
 In order to check whether homogeneity is present 
 in a given sample one has to use 
 the conditional density $\Gamma(r)$ 
 defined as \cite{pie87}
 \be
 \label{eq1}
 \Gamma(r) = \frac{\langle n(r_*) n(r_*+r) \rangle}{\langle n 
 \rangle} =
 \frac{AD}{4 \pi} r^{D-3}
 \ee
 where the last equality holds in the case of a fractal
 distribution
 with dimension $D$ and prefactor $A$ (we follow here  
 the same definitions of \cite{guzzo}). In the case of an 
 homogenous distribution ($D=3$) the conditional density 
 equals the average density in the sample. 
 Hence the conditional density is the suitable 
 statistical tool to identify 
 fractal properties (i.e. power law correlations
 with codimension $\gamma=3-D$) as well as 
 homogeneous ones (constant density with sample size).
 If there exists a transition scale  $\lambda_0$ towards homogenization,
 we should find $\Gamma(r)$ constant for scales 
 $r \gtapprox \lambda_0$.

 It is simple to show that 
 in the case of a fractal distribution the 
 usual $\xi(r)$ function in a spherical sample of radius $R_s$ 
 is \cite{pie87} \cite{cp92}
 \be
 \label{eq2}
 \xi(r) = \frac{D}{3} \left( 
 \frac{r}{R_s} \right)^{D-3} -1 \; .
 \ee
 From Eq.\ref{eq2} we can see two main problems of the 
 $\xi(r)$ function: its amplitude depends on the sample size $R_s$
 (and the so-called correlation length $r_0$, defined as 
 $\xi(r_0) \equiv 1$, linearly depends on $R_s$) 
 and it {\it has not} a power law behavior.
 Rather the power law behavior is present 
 only at scales $r  \ll r_0$, and 
 then it is followed by a sharp break in the log-log plot
 as soon as $\xi(r) \ltapprox 1$. Such a behavior does not
 correspond  to any real change of the correlation 
 properties of the 
 system (that is scale - invariant by definition) and it 
 makes extremely difficult the estimation
 of the correct fractal dimension as it shown in Fig.\ref{fig1}. 
 In particular if the sample size is not large enough 
 with respect to the 
 actual value of $r_0$, 
 the codimension estimated by the $\xi(r)$ function 
 ($\gamma \approx 1.7$) 
 is systematically larger than $3-D$ ($\gamma \approx 1$) 
 \cite{slmp97}.
 \bef
 \epsfxsize 8cm
 \centerline{\epsfbox{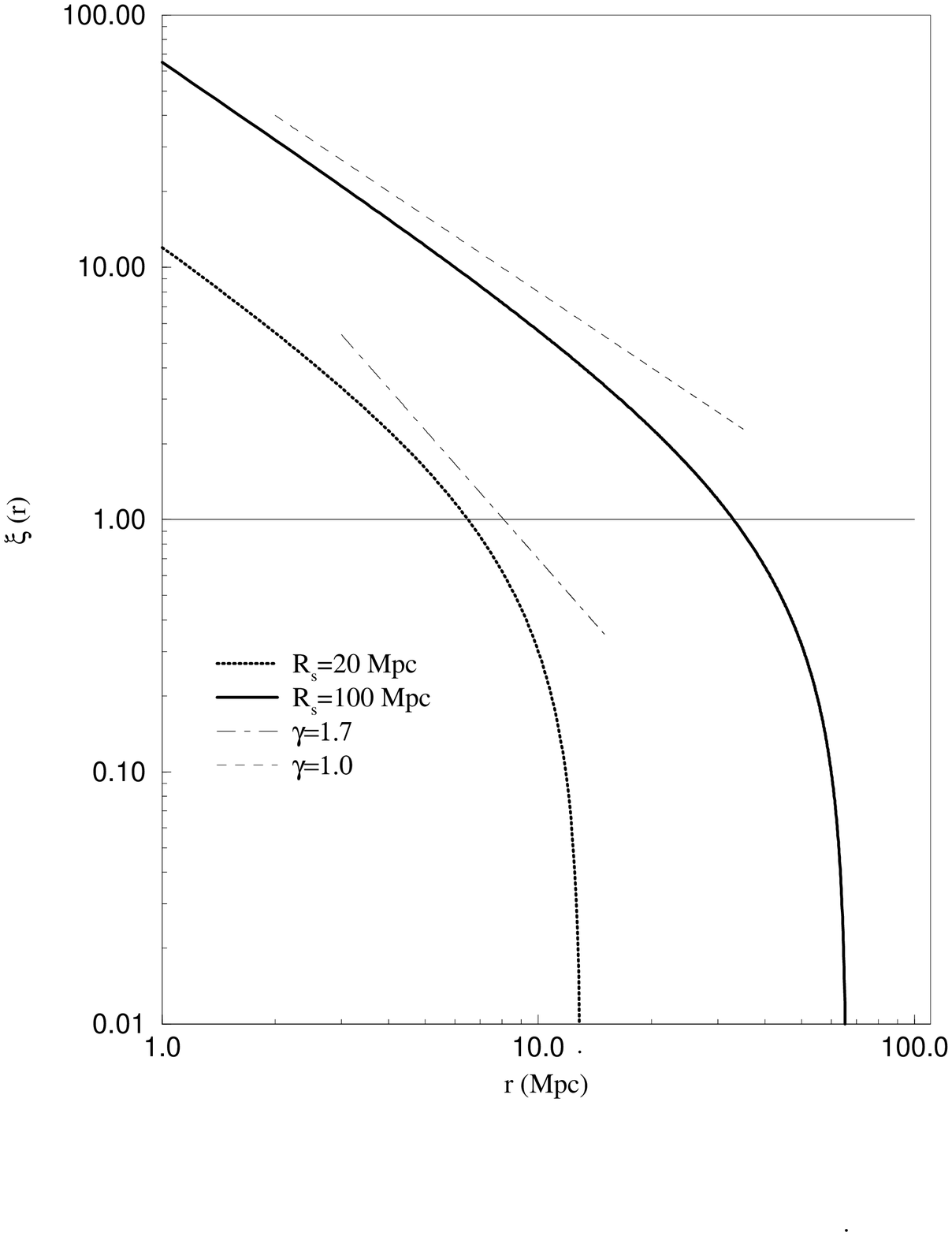}} 
 \caption{\label{fig1} The $\xi(r)$ function computed 
 for a fractal with dimension $D=2$ ($\gamma=3-D=1$)
 in two samples with different depth $R_s$. In the first case 
 $R_s=20 Mpc$
 one obtains an higher value for the correlation exponent 
 $\gamma=1.7$
 because one performs the fit in the region of length scales 
 $r \sim r_0$, as the power law behavior is poorly defined.
 In the second case the power law behavior of
 $\xi(r)$
 is more extended ($R_s=100 Mpc$)
 and it is now possible to fit in the region 
 $r \ll r_0$. Here one obtains the correct value for
 the correlation exponent $\gamma=1$. 
 However the effective 
 depth of the various galaxy samples
 is in the range 
 $\sim 10 \div 30 \hmp$ 
 (see Table 1) and hence this effect is very 
 important
 in the real data.} 
 \eef

 Given this situation it is clear that the $\xi(r)$ analysis 
 is not suitable
 to be applied unless
 a clear cut-off towards homogenization 
 is present in the samples analyzed. As this is not the case, 
 as also
 Dr. Guzzo stressed, it is appropriate and convenient to use 
 $\Gamma(r)$ instead of $\xi(r)$. We have discussed 
 in detail in \cite{slmp97} that the use of the 
 correct statistical methods is complementary 
 to a change of perspective from a theoretical
 point of view.


 \section{Galaxy distribution at scale 
 $r \le 100 \div 200 \hmp$ }

 From the previous discussion 
 it seems that Dr. Guzzo agrees with us about
 the fact that, unless a well defined value of the average 
 density has been established, and found to be independent 
 on the sample size, the usual $\xi(r)$ analysis fails.
 Hence we should focus on the determination of the average density
 rather than on $r_0$: the latter quantity being  meaningful
 unless
 an homogeneous distribution has been found. 

 The first test on fractal versus homogeneous
 properties concerns the relation between
 the sample size $R_s$ and the so-called
 correlation length $r_0$ measured 
 in redshift catalogs. The conclusion of
 Dr. Guzzo is that this relation is not the linear
 one predicted in the case of fractal distributions.
 Here we briefly report our analysis 
 of the $r_0-R_s$ relation, which 
 significantly disagrees with the one of
 Dr. Guzzo.

In Tab.\ref{tab1} we report the characteristics of the 
various catalogs we have analyzed by using
 the methods illustrated in Sec.2.
\begin{table}
\caption{\label{tab1}The volume 
limited catalogues are characterized by the following 
parameters:
- $R_d (\hmp)$ is the depth of the catalogue
- $\Omega$ is the solid angle
- $R_s (\hmp)$ is the radius of the largest sphere 
that can be contained in the catalogue volume. 
This gives the limit of statistical validity of the sample.
- $r_0(\hmp)$  is the length at which $\xi(r) \equiv 1$.
- $\lambda_0$ is the eventual real crossover to a homogeneous 
distribution that is actually never observed. 
The value of $r_0$ 
is the one obtained 
in the deepest VL  sample. 
 The CfA2 and SSRS2 
 data are not yet available and in this case the values
 of $r_0$ have been taken by. Park \etal 1994 
 and Benoist \etal 1996 respectively.
(Distance are expressed in $\hmp$).
}
\begin{tabular}{|c|c|c|c|c|c|c|}
\hline
     &      &          &    &              &  &       \\
\rm{Sample} & $\Omega$ ($sr$) & $R_d  $ & $R_s  $ 
& $r_0  $& $D$ & $\lambda_0 $ \\
     &       &    &    &    &              &       \\
\hline
CfA1         & 1.83      & 80 & 20 & 6  & $1.7 \pm 0.2$ & $>80$    \\
CfA2         & 1.23      & 130& 30 & 10 & $\sim 2.0   $ & $ ? $    \\
PP           & 0.9       & 130& 30 & 10 & $2.0 \pm 0.1$ & $> 130$  \\
SSRS1        & 1.75      & 120& 35 & 12 & $2.0 \pm 0.1$ & $ >120$      \\
SSRS2        & 1.13      & 150& 50 & 15 & $\sim 2.0   $ & $?$      \\
Stromlo-APM  & 1.3       & 100& 35 & 12 & $2.2 \pm 0.1$ & $ > 150$      \\
LEDA         & $\sim 5 $ & 300& 150& 45 & $2.1 \pm 0.2$ & $>150$   \\
LCRS         & 0.12      & 500& 18 & 6  & $1.8 \pm 0.2$ & $> 500$    \\
IRAS$2 Jy$   & $\sim 5$  & 60 & 20 & 5  & $2.0 \pm 0.1$ & $ > 50$\\
IRAS$1.2 Jy$ & $\sim 5$  & 80 & 30 & 8  & $2.0 \pm 0.1$ & $> 50$\\
ESP          & 0.006     & 700& 8  & 3  & $1.9 \pm 0.2$ & $>700$   \\
              &          &         &       &      &  &       \\
\hline
\end{tabular}
\end{table}
 Here we consider in more detail only three catalogs, where 
 our estimations of $r_0$ and/or $R_s$ significantly disagree 
 with those of Guzzo and a 
  detailed explanation of the analyses 
 of the catalogs shown in Tab.1 can be found in \cite{slmp97}.
However before doing this we would like to stress 
two important points:

{\bf 1.} Given a certain sample of solid angle $\Omega$ and depth $R_d$,
it is important to define which is 
 the maximum distance up to which it 
is possible to compute the correlation function ($\Gamma(r)$ or $\xi(r)$). 
As discussed in \cite{cp92},
 we have limited our analysis to an
effective  depth
$R_{s}$ that is of the order of the radius of the maximum
sphere fully contained in the sample volume.
In such a way we eliminate from the statistics
the points for which a sphere of radius {\it r} is not
fully included within the sample boundaries.
Hence we do not make any assumption on the treatment of
the boundaries conditions.
Of course, by
doing this, we have a smaller number of points
and we stop our analysis  at a  smaller depth than that
of other authors.

 The reason why
$\Gamma(r)$ (or $\xi(r)$) cannot
be computed for $r > R_{s}$
is essentially the following.
 When one evaluates the correlation
function
(or the power spectrum \cite{sla96}) beyond $R_{s}$,
then one  makes explicit assumptions on what
lies beyond the sample's boundary.  In fact, even in absence of
corrections for selection effects, one
is forced to consider incomplete shells
calculating $\Gamma(r)$ for $r>R_{s}$,
thereby
implicitly assuming that what it is not seen in the part of the
shell not included in the sample is equal to what is inside (or other
similar weighting schemes).
In other words, the standard calculation
introduces a spurious homogenization which we are trying to remove
\cite{cp92} \cite{slmp97}.

We have done  a test \cite{slmp97} 
on the homogenization effects
 of the incomplete shells on artificial distributions as well as on
real catalogs, finding that the flattening of the 
conditional density
is indeed introduced owing to  the weighting,
and does not correspond to any real feature in the galaxy distribution.
These results differ from those of \cite{pro94} (this is
reported in the Appendix A in Dr. Guzzo's paper)
probably because they did not take into account
finite size effects in the generation of artificial samples and
they considered ensemble averages of the conditional density 
(see \cite{slmp97} for a more detailed discussion).

{\bf 2.} We do not use  weighting schemes, and hence 
our analysis concerns only {\it volume limited } (VL)
samples. The use of magnitude limited (ML) samples and the
weighting schemes inevitably requires
a-priori assumptions on nature of the distribution
\cite{slmp97}.

Now we discuss in detail our disagreement with Tab.1 of the 
paper of Guzzo.

\begin{itemize}

\item ESP. In this case  the estimation of $R_s$ slightly differs
from that of Guzzo, probably because we have not used 
the relativistic corrections (\cite{slmp97} see below). Also
the value of $r_0$ is slightly different ($r_0 \sim 3 \hmp$
 instead of
the measured $r_0 \sim 4.5 \hmp$), and this is probably due 
to the fact that ESP does not cover a continue solid angle in the 
sky, as it is a collection of pencil beams. Such a 
situation necessarily requires the introduction of 
spurious treatments of the boundary conditions (see point {\bf 1}).

\item LCRS. This survey has the peculiar property of being 
limited by two limits in apparent magnitude (a lower and 
an upper one). In order to construct a VL sample
in this case,
 one has to impose two limits in distances and
correspondingly two in absolute magnitude. This
is the origin of a smaller $R_s$ 
in our Table 1 than this reported by Guzzo ($R_s \sim 32 \hmp$)
\cite{slmp97}. This implies a smaller $r_0$,
 much closer to the measured one.

\item Stromlo/APM. We have extensively analyzed this catalog 
in \cite{slmp97} and \cite{slm97} and the value of
$r_0$ is reported in Tab.1. 
Due to the sparse sampling strategy 
adopted to construct this catalog,
we are able to measure the correlation properties up to
$R_s \sim 40 \hmp$ and not $83 \hmp$ as reported by Guzzo.
The disagreement with the work 
of Loveday \etal \cite{lov96} ($r_0 \approx 12 \hmp$ rather than
$r_0 \approx 5 \hmp$) is probably due to the treatment of
the boundary conditions and their use of ML samples rather than 
VL ones (i.e. they used weighting schemes with 
the luminosity selection function). In any case, we stress again,
the proper test is check whether the conditional density has 
a power behavior \cite{slm97}.
\end{itemize}

 We show in Fig.\ref{fig80} the results
 of the conditional density
 determinations in various redshift surveys \cite{slmp97} \cite{mslgpa97}.
 All the available data 
 are consistent with each other and show fractal correlations with
 dimension $D = 2.0 \pm 0.2$ up to the deepest scale probed up to
 now by the available redshift surveys, i.e. $\sim 150 \hmp$.
 A similar result has been obtained by the analysis of galaxy cluster
 catalogs \cite{msla97} \cite{slmp97}.
 \bef 
 \epsfxsize 10cm 
 \centerline{\epsfbox{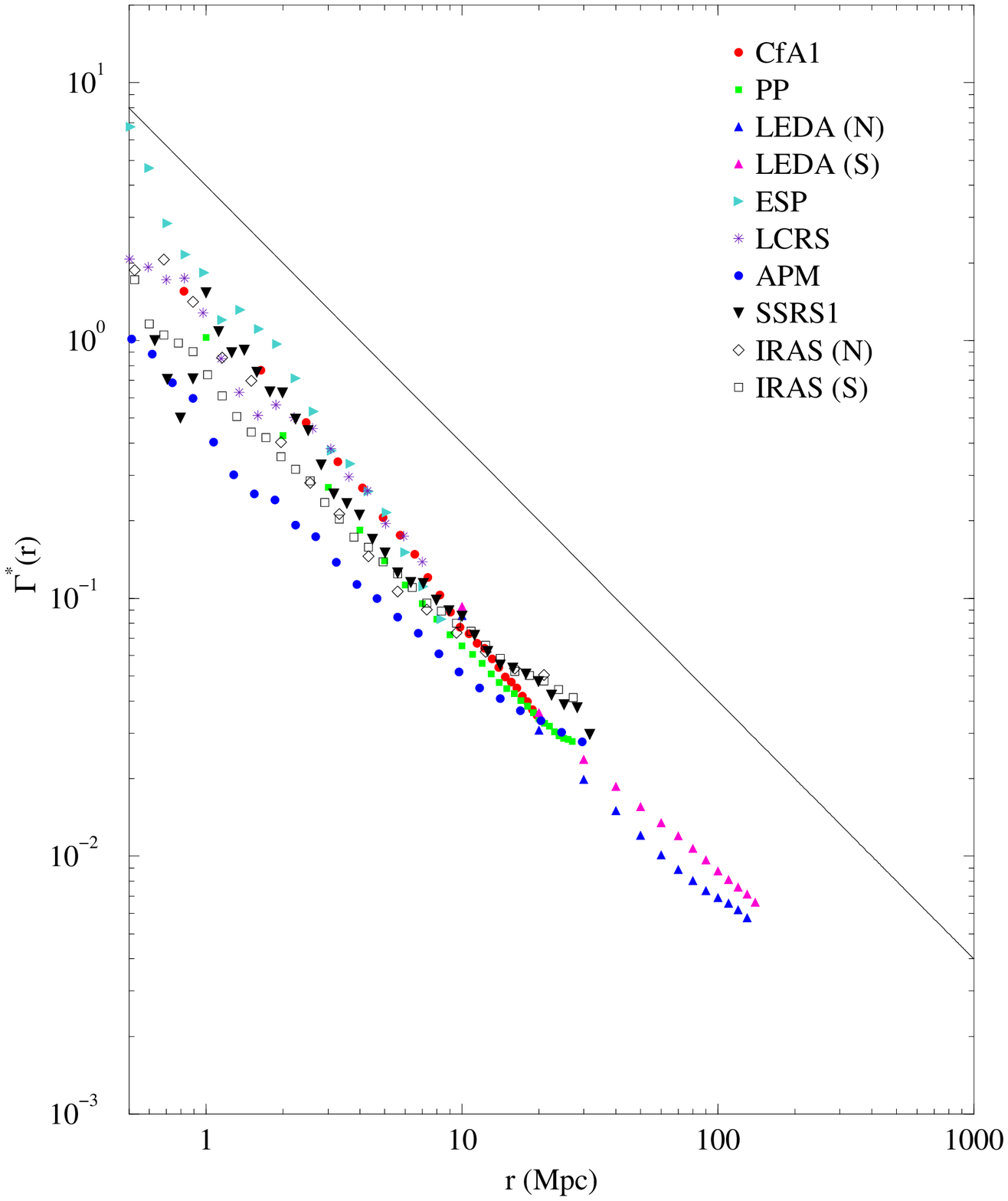}}  
 \caption{\label{fig80} 
 The spatial density   $\Gamma(r)$     computed in some 
 VL samples of CfA1, PP, LEDA, APM, ESP, LCRS, SSRS1, IRAS 
 and ESP (Sylos Labini \etal, 1997). } 
 \eef 

 Finally it must be noted that
 Dr. Guzzo
 based a part of 
 his arguments on the fact that a luminosity bias is responsible
 of the shift of $r_0$ with sample size (see also \cite{dav88}
 \cite{dav97}).
 We do not enter into a detailed explanation of the inconsistencies of 
 the usual argument about luminosity segregation \cite{dav88} \cite{par94}
 \cite{ben96} and we refer to \cite{slmp97} for a more detailed analysis.
 However we would like to remark that  
 the authors (e.g.  \cite{par94} \cite{ben96}) 
 who have addressed  this 
 concept  have  never presented {\it any quantitative argument
 that explains the shift of $r_0$ with sample size.}
 In this respect, an exception is represented by the 
 paper of Davis \etal (1988) 
 in which the authors claim that
 $r_0$ behaves as $R_s^{0.5}$. However for what concerns
 luminosity segregation, the meaningful parameter must be the absolute 
 magnitude limit of the volume limited (VL)
 sample considered rather than the depth ($R_s$).
 Having fixed the limiting apparent magnitude of the 
 catalog, at each $R_s$ would correspond a well defined 
 absolute magnitude. The brightest galaxies are  present yet
 in samples like CfA1, and hence according to
 the luminosity segregation paradigm, there is no reason  one should 
 expect that in deeper sample (like CfA2 or SSRS2) $r_0$
 is increased. However this is actually the case \cite{ben96} \cite{par94}.

 We have discussed in detail in \cite{slp96} 
 that the observation
 that the giant galaxies are more clustered
 than the dwarf ones, i.e. that the massive elliptical
 galaxies lie 
 in the peaks of the density field, is a consequence of the
 self-similar behavior of the whole {\it matter distribution}. The increasing
 of the correlation length of the $\xi(r)$ has
 nothing to do with this effect, rather it is
related to the sample size.

 A last point requires a further clarification. Dr. Guzzo
 claims that $\Gamma(r)$  is a power law with $D \sim 1.2$ up to
 $ r \sim 3.5 \hmp$, then it shows  a different scaling  
 between $\sim 3\hmp$ and $30 \hmp$ with fractal 
 dimension $D\approx 2$
 \cite{guz91}. In our view there is a subtle point in this discussion
 that has not been previously appreciated.

Suppose now, for simplicity,  we have a 
spherical sample of volume $V$ in which there 
are $N$ points, and we want to measure the conditional density.
It is possible to compute  the average  distance 
between neighbor galaxies $\langle \Lambda \rangle$, in a fractal
distribution with dimension $D$, and the result is 
\be
\label{e5}
\langle \Lambda  \rangle = \left(\frac{1}{A}\right)^{\frac{1}{D}} \Gamma
\left(1 + \frac{1}{D} \right)
\ee
where $\Gamma$ is the Euler's gamma-function \cite{sl97b}.
(Note that the prefactor $A$ is dependent on the luminosity
selection function of the VL chosen).
Clearly this quantity   is related 
to the lower cut-off of the distribution $A$ (eq.\ref{eq1}) 
and to the fractal dimension $D$.
If we measure the conditional density at distances 
$ r \ltapprox \langle \Lambda  \rangle$, 
we are affected 
by a {\it finite size effect}. In fact, due the depletion of points at these 
distances 
we underestimate the real conditional density finding an higher value 
for the correlation exponent (and hence a lower value for the fractal 
dimension). 
In the limiting case at the distances $ r \ll \langle \Lambda  \rangle $, 
we can find almost no points and 
the slope is
$\gamma=-3$ ($D=0$). 
In general, when one  measures $\Gamma(r)$ at distances 
which correspond to 
a fraction of $ \langle \Lambda  \rangle$, 
one finds systematically an higher value of the 
conditional density exponent. 
Such a trend  is completely spurious and due to the depletion of
points at such distances. It is worth to notice that this effect
gives rise to a curved behavior of 
$\Gamma^*(r)$  (the integral of $\Gamma(r)$ see \cite{cp92} \cite{slmp97})
 at small distances, because 
of its integral nature. This is exactly the case of the deepest VL 
of Perseus-Pisces which Guzzo \etal (1991) considered in their analysis,
and for which $ \langle \Lambda  \rangle  \sim 8 \hmp$. A clarifying test 
in this respect would be to check whether this change of slope
is actually present also in the others VL samples of the same survey, 
which have a larger number of points (and hence a lower $\langle \Lambda 
 \rangle$).
This test has been performed by \cite{slmp97} and the conclusion is
that the change of slope is {\it due a finite size effect} rather
being an intrinsic property of galaxy distribution.

Finally we would like to point out another point that 
is inconsistent in the arguments of Dr. Guzzo. He and his collaborators
\cite{guz91} found that $\Gamma(r)$ flattens at $ \gtapprox 30 \hmp$. 
Up to this distance $r_0$ must be a linear fraction of $R_s$ in view
of eq.\ref{eq2} and does not depend to any luminosity bias !
 In our opinion \cite{slmp96} \cite{slmp97}, 
 this flattening 
 is due to an incorrect treatment of the boundary conditions
(see point {\bf 1}). However if this behavior is real,
 Dr. Guzzo should conclude that galaxy distribution
 is homogeneous at 
$\sim 30 \hmp$ and 
not at $\sim 100 \div 200 \hmp$ as he claims.

We conclude this discussion pointing out 
another element: the analysis of LEDA. Dr. Guzzo
claims that the results coming from LEDA \cite{dinella} \cite{amendola} are 
``as having no meaning whatsoever''. In our opinion 
such a {\it strong conclusion} must be supported by quantitative arguments.
We have done several tests on the LEDA database \cite{dinella}
\cite{amendola} \cite{slmp97} and we have concluded that,
although this sample is highly incomplete, the statistical results
obtained are rather stable and robust. We refer to the previously
mentioned papers for a more specific discussion.


\section{Galaxy distribution at scale $r \ge 100 \div 200 \hmp$}

We discuss now the behavior of the radial density in general, 
and then we consider the two cases of PP and ESP (see \cite{slgmp96}
\cite{mslgpa97} and \cite{slmp97} for a more detailed discussion
on this point).

We focus  on the possibility of
extending the sample effective depth $R_{s}$.
 In order to
discuss this question,
 it is important to analyze the properties
 of the {\it  small scale
fluctuations}.  
To this aim, we introduce the conditional density in 
the 
volume $V(r)$ (that can be a portion of a sphere) 
as {\it observed from the
 origin}, defined as 
 \be 
\label{new1} 
\Gamma_i^{VL} (r) = \frac{N_i^{VL}(<r)}{V(r)} = \frac{3Ap}{4\pi} r^{D-3}  \; .
\ee  
where the factor $p$ comes from the fact that  a VL sample
contains only a fraction
$N_i^{VL}(R) = p \cdot N(<R)$  (where $p<1$)
 of the total number $N(<R)$ of galaxies in $V(R)$. 
If $\phi(L)dL$ is
 the fraction of galaxies whose absolute luminosity ($L$) is between 
$L$ and
$L+dL$, $p$ is given: 
\begin{equation}
\label{p1}
 0< p = \frac{\int_{L_{VL}}^\infty  \phi(L)dL} 
{\int^\infty_{L_{min}} \phi(L)dL} < 1
\end{equation}
 In Eq.\ref{p1}
$L_{VL}$ is the minimal absolute luminosity that 
characterizes the VL
sample and $L_{min}$ is 
the fainter absolute luminosity (or magnitude $M_{min}$)
surveyed in the catalog (usually $M_{min} \sim -11 $).
 Computing $\Gamma_i(r)$, we expect (Fig.\ref{fig61}) 
not to see any galaxy up to a certain 
distance $\langle \Lambda \rangle$.  
For distances somewhat 
larger than
$\langle \Lambda \rangle$, we expect therefore a raise of 
the conditional density 
because we
are beginning to count some galaxies and 
$\Gamma_i(r)$ is affected by the fluctuations due to the 
low statistics. 

It is therefore important to be able to
estimate and control the {\it minimal
statistical length} $\lambda$, which separates 
the fluctuations due to the low statistics 
from the genuine behavior  of the distribution. 
A simple argument for the 
determination on the length $\lambda$
is the following \cite{slgmp96} \cite{mslgpa97} \cite{slmp97}.
At small scale, where there is a small number of galaxies, 
there is an additional 
term, due to shot noise, superimposed to the power
law behavior of $\Gamma_i(r)$,
 that destroys the genuine correlations 
of the system.( As we have discussed in \cite{slmp97} 
there should be considered also an {\it 
intrinsic oscillating term}
for non-averaged out quantities: however for
sake of clarity we avoid this discussion here.)
  Such a fluctuating term can be 
erased out by making an average over all the points 
in the survey. 
On the contrary, in the observation from one point, 
 when the number of galaxies is large enough the shot noise
 becomes negligible. Roughly, this happens when
 the number of points is lager than 
 $\sim 30$ (see Fig.\ref{fig61}).
\bef 
\epsfxsize 12cm 
\centerline{\epsfbox{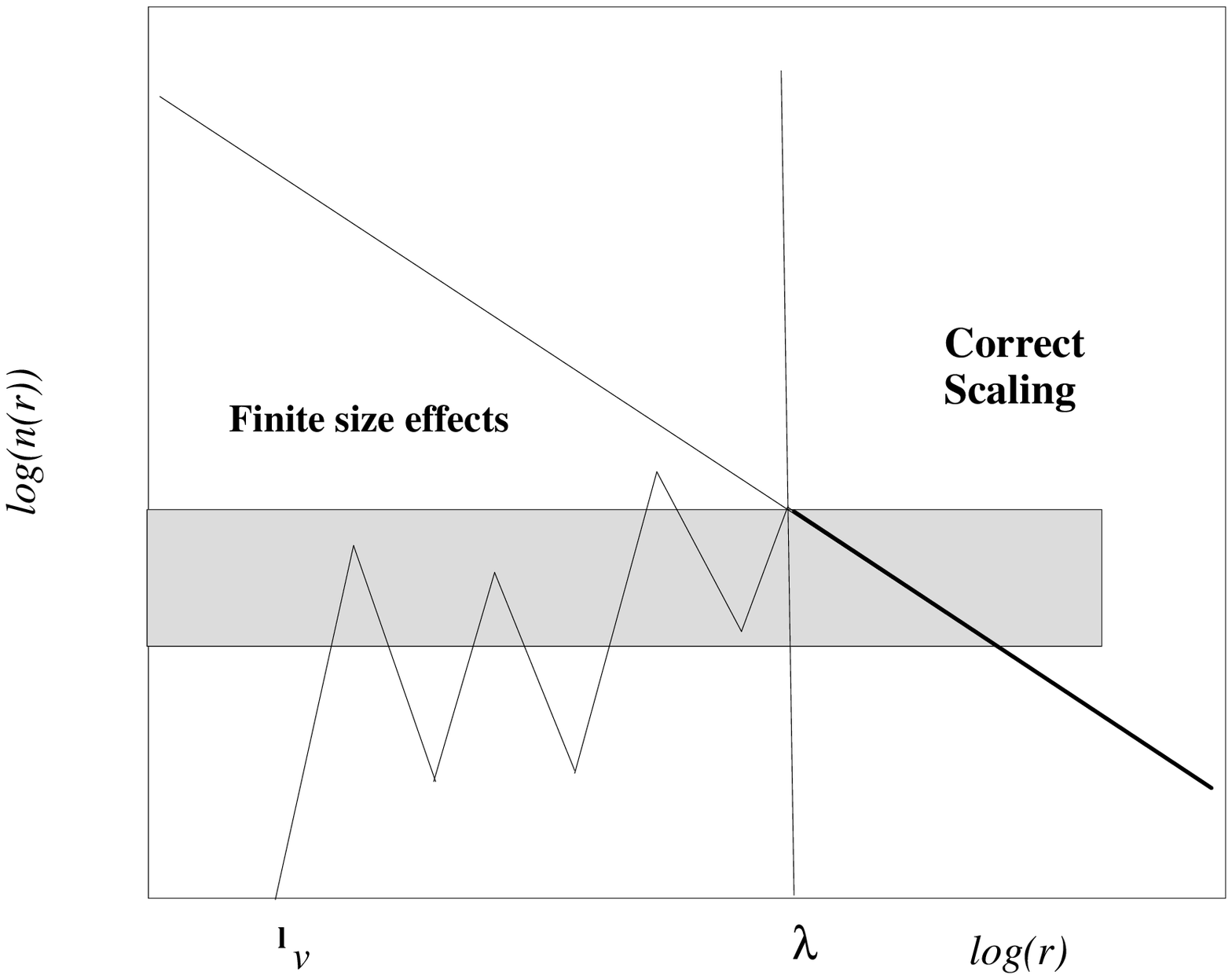}} 
\caption{\label{fig61}  Behavior of the density computed from one
 point, in the case of a fractal
 distribution. At small distances
 below the average mean separation between neighbor galaxies,
 one finds no galaxies. 
 Then the number of galaxies starts to grow, but this regime
 is strongly affected by finite size fluctuations. Finally the
 correct scaling region $r \approx \lambda$ is reached. In the
 intermediate region the density can be approximated roughly by a
 constant value.   This leads to an apparent exponent $D \approx
 3$ 
for the integrated counts.
This exponent is not real but just due to the size effects.
 } 
 \eef 
 This condition gives (from Eq.\ref{new1})
\begin{equation}
\label{v3}
\lambda =  5 \left( \frac{4 \pi}{A p \Omega} \right)^{\frac{1}{D}} \approx 
\frac{20 \div 60 h^{-1} Mpc}{\Omega^{\frac{1}{D}}}. 
\end{equation}
 for a typical VL sample with $M_{VL} \approx M^*$, where 
 $A$
 corresponds to the amplitude of the conditional density of all
 galaxies \cite{slgmp96} \cite{slmp97}. 
 This can be estimated from the amplitude of $\Gamma(r)$ 
 in a VL sample 
 divided by the correspondent $p$ as
 defined in Eq.\ref{p1}.  We find  (for typical catalogues) 
 $B \approx 10\div 15 (h^{-1} Mpc)^{-D}$ \cite{slgmp96}.
 The corresponding value of $\lambda$ are reported in Tab.2.
\begin{table} \begin{center} 
\begin{tabular}{|c|c|c|} 
\hline         
&       &     \\ 
Survey & $\Omega (sr)$ & $\lambda (\hmp)$  \\ 
   &    &     \\ 
\hline 
CfA1            & 1.8      & 15 \\  
CfA2 (North)& 1.3      & 20  \\        
SSRS1          & 1.75    & 15  \\    
SSRS2          & 1.13    & 20 \\  
PP                & 1         & 40  \\
LEDA            & 2 $\pi$& 10 \\      
IRAS            & 2 $\pi$ & 15 \\        
ESP             &   0.006  & 300  \\        
\hline
\end{tabular} 
\caption{The {\em minimal statistical length 
\label{tablambda}} $\lambda$ for several redshift surveys.}
\end{center} \end{table}

The difference between $\Gamma(r)$ and $\Gamma_i(r)$
is straightforward: the latter quantity is not
an average one. However it is an integrated one.
The differential density is shells is clearly much more noisy,
because it is not averaged neither integrated.
We have computed the $\Gamma_i(r)$ in the various VL samples 
of Perseus-Pisces redshift survey, 
and we show the results   in Fig.\ref{fig62}.
 In the less deeper VL samples (VL70, VL90) 
 the  smooth behavior of the density
 is weakly defined because in this case the
 finite size effects  are very important
 as the distances involved are 
 $r < \lambda $ (Eq.\ref{v3}). We note that 
 about  the same scales we
 find a very well defined power law behavior by the
 correlation  function analysis.
 In the deeper VL samples (VL110, VL130)  a smooth
 and well defined behavior is reached for distances larger 
 than the scaling distance ($\Omega =0.9 \, sr$)
 $r \approx \lambda \sim 50 h^{-1}Mpc$. The fractal dimension is
  $D \approx 2$ as the one measured by 
 the average conditional density.

 \bef 
 \epsfxsize 8cm 
 \centerline{\epsfbox{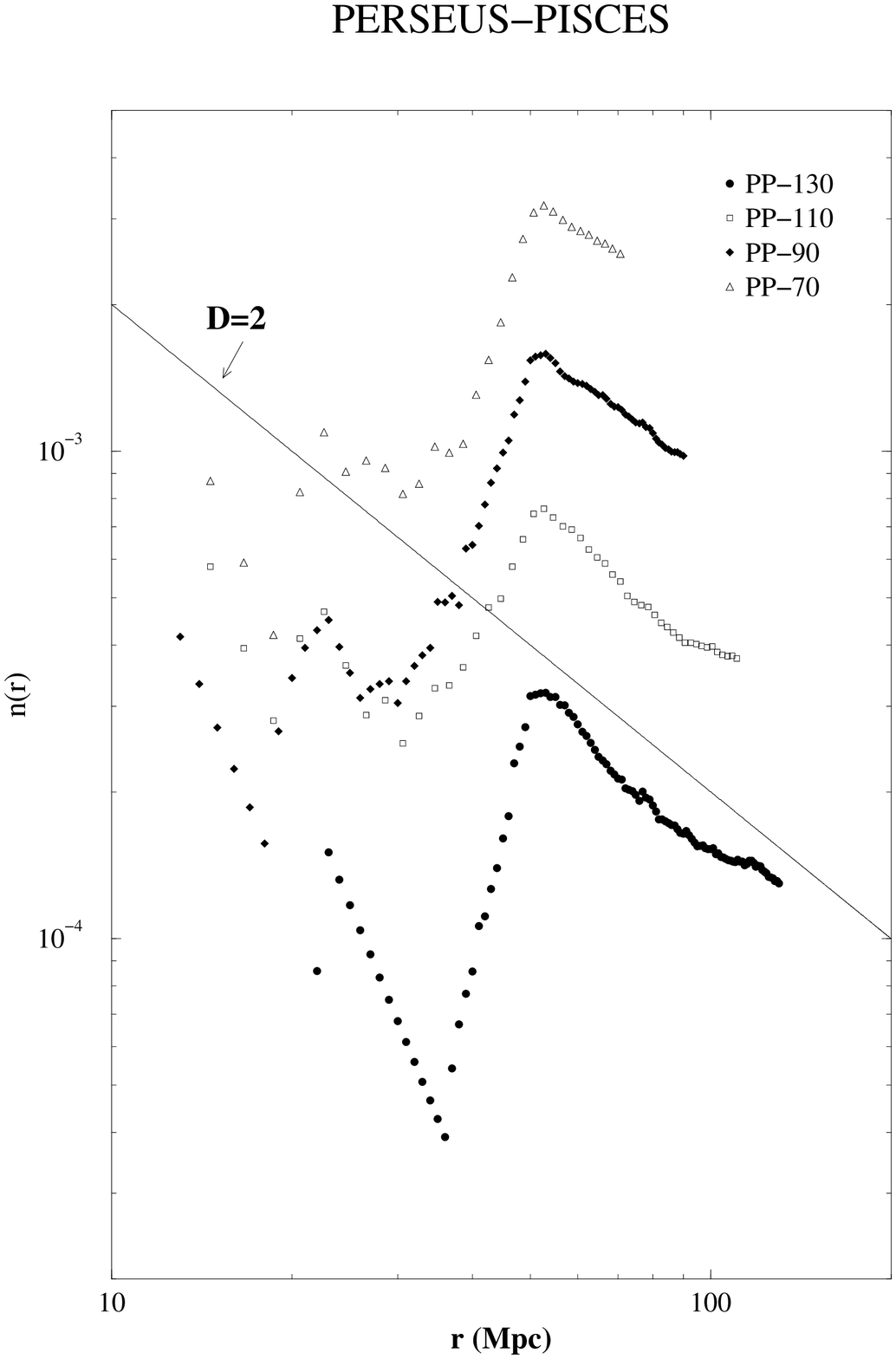}} 
 \caption{ The spatial density 
 $\Gamma_i(r)$ computed in the VL sample
 cut at $70, 90, 110, 130  h^{-1}Mpc$ . In the case of VL70 
 and VL90  the
 density is dominated by large fluctuations and it has not reached
 the scaling regime. In the samples  VL110   and VL130 the density 
 is
 dominated by large fluctuations only at small distances, while at
 larger distances, after the Perseus Pisces chain at $50
 h^{-1}Mpc$, a very  well defined power law behavior is shown,
 with
 the same exponent of  the correlation function  (i.e. $D=2$)
 \label{fig62}}
 \eef

 For relatively small volumes it is possible to
 recover the correct scaling behavior for scales of order of
 $\langle \Lambda \rangle$
 (instead of $ \sim 10 \langle \Lambda \rangle$)
 by averaging over several
 samples or, as it happens in real cases, over several points of
 the same sample when this is possible. Indeed, when we compute
 the
 correlation function, we perform an average over all the points of
 the system even if the VL sample is not deep enough to satisfy
 the
 condition expressed by Eq.\ref{v3}. 

 Now, if we look at the density in shells, 
 this will be dominated by fluctuations, as it is not a
 cumulative distribution. Note that if the distribution
 would became homogenous, as Guzzo \etal  \cite{guz91} found,
 at $\sim 30 \hmp$, then  at 3 times the homogeneity scale the 
 distribution would present a rather regular behavior.
 This is clearly not the case.

 Let us now briefly discuss the properties of the ESP catalog.
 As the redshifts involved are quite large a particular care is
 devoted to the construction of the VL subsamples for 
 the case of ESP. We have used
 various {\it distance-redshift} ($d(z)$) and {\it
 magnitude-redshift} ($m(z)$) relations.
 Moreover  the data are selected in the blue-green
 and even if the redshifts of galaxies are moderate ($z\le0.2$),
 K-corrections are needed to compute the absolute magnitude of
 galaxies. The corresponding {\it magnitude-redshift}
 relation is
 \be
 \label{eqmlesp}
 m-M=25+5\log_{10}\left(d(z,q_0)(1+z)\right) + a(T)z
 \ee
 where we have used the functional forms of the K-correction
 $a(T)z$ as a function of redshift. $a(T)$
 depends on the morphological type $T$ and goes from $a \sim 2$
 for the {\it Scd } galaxies to $a \sim 3.7$ for the {\it E/S0}
 galaxies.
 It is not possible to apply the K-correction to each
 morphological type because over the $\:17^{th}$ magnitude it is
 not possible to recognize the Hubble type from visual inspection.
 To overcame this  problem  we have adopted
 a statistical approach: we have assumed 
 various percentage of late 
 and early-type galaxies. The
 observed percentage in nearer samples is $\:70\%$ of late and
 $\:30\%$ of early type  galaxies. We have performed a number of
 tests by varying these percentages to show that the final result
 depends weakly on the adopted values. We note that varying these
 percentages change the number of points in the VL 
 subsamples as the absolute magnitude can change of about a unit.

The behavior of the {\em number-distance relation} in the standard
 FRW model depends strongly on the value of the deceleration
 parameter $\:q_{0}$ for high redshift $\:z>0.5$, while for $\:z
 <0.2$ the relativistic corrections are very small for all
 reasonable $\:q_{0}$.

 All the samples show a 
 highly fluctuating behavior for the space density 
 as it is reported by Dr. Guzzo.
 It is very difficult in this case to identify a 
 clear power law behavior. However we   stress
 that an homogeneous distribution (say at $ 50 \hmp$)
 would show a very smooth and flat behavior for 
 $\Gamma_i(r)$ at these distance scales. This is clearly not the 
 case.

 It is very interesting, in our opinion, to study also the case of
 a purely Euclidean space. In this case we can write the $d(z)$
 relation simply as: 
 \be 
 d=\frac{c}{H_0}z
 \ee 
 and the $m(z)$ relation (without K-corrections): 
 \be 
 m-M=5\log_{10} \left( d(z)  \right) +  25 \; .
 \ee
 In this case the volume grows as $\sim z^3 \sim d^3$.
 The  behavior of the radial density in the Euclidean case is 
 reported in Fig.\ref{fig69}.
 \bef 
 \epsfxsize 8cm 
 \centerline{\epsfbox{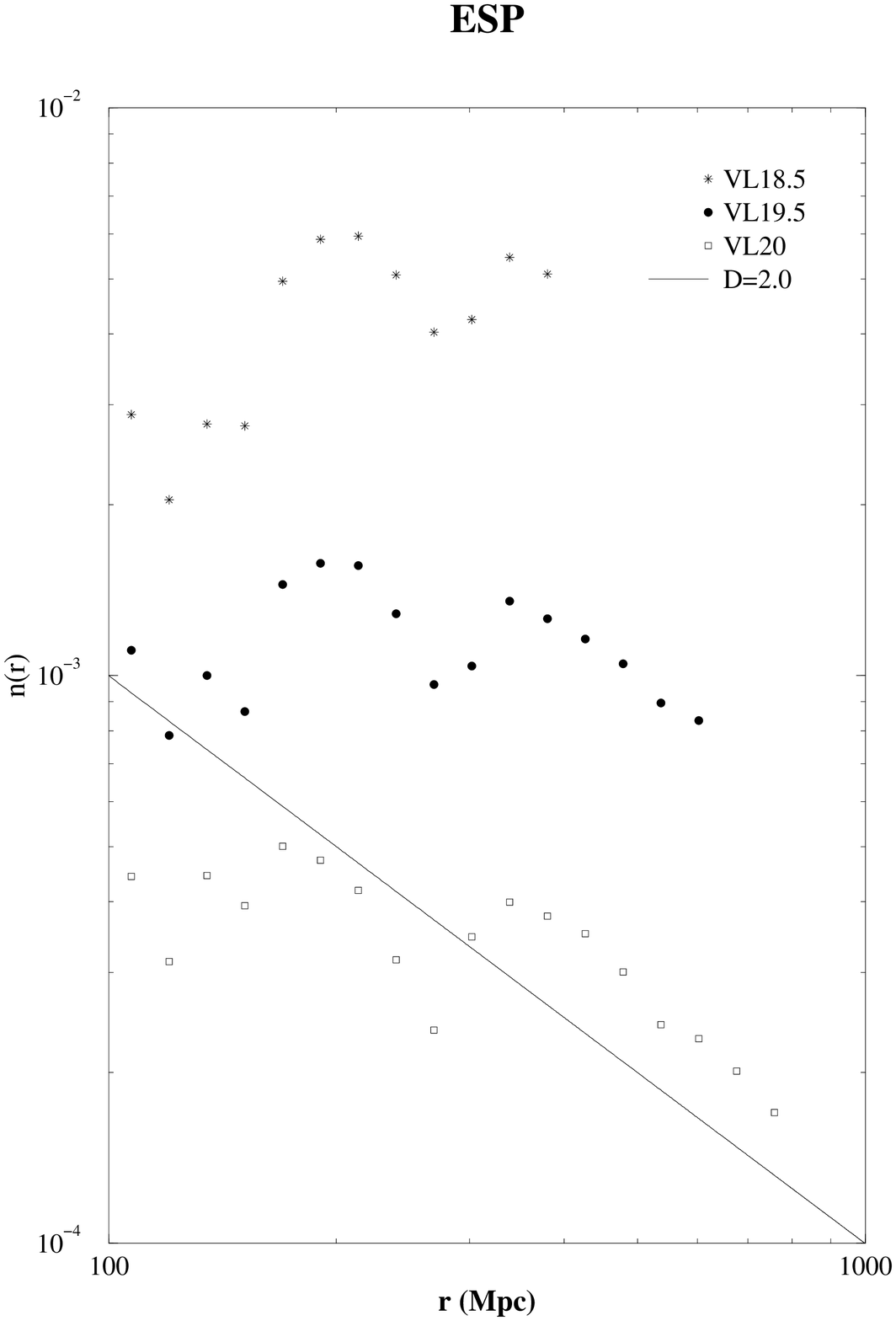}} 
 \caption{\label{fig69} The spatial density   $\Gamma_i(r)$ 
 computed
 in a VL samples of ESP. A the power law behavior
 with $D \approx 2$  is shown for $r \gtapprox 300 \hmp \approx
 \lambda$, even though with strong fluctuations}
 \eef 
 From  this figure it
 is possible to see  the effect of the finite size
 fluctuations: the scaling region begins at 
$\lambda \sim 300 \div 320 h^ {-1}Mpc$ for all the VL subsamples. For
 smaller distances the statistical fluctuations dominate the behavior
 of $\Gamma_i(r)$. 
 It seems that in this case the power law behavior 
 for $r \gtapprox \lambda \approx 320 \hmp$
 is better defined than in the previous case.
 The fractal dimension turns out to be $D \approx 2.2$.

 The {\it K-correction}, that must be applied for
 the Hubble shift of the galaxy spectra, can change 
 the absolute magnitude of a unit. This
 correction is due to 
 a systematic effect for each morphological type. As
  we put these correction in a random way we are mixing a
 systematic effect with a random correction. In this way we can
 only {\it  check the stability of the results} but we cannot hope
 to obtain a better fit \cite{slmp97}
 By doing this \cite{slmp97} we find a marginal power law decaying
 of the conditional density.
 Therefore we may conclude that
there is a weak evidence that 
the fractal dimension is $\:D \approx 2$ in this sample
due to the poor statistics. 

There are several other evidences in our opinion that point
towards a fractal distribution of galaxies at very large scale, 
and in particular they are the density behavior from one point
in the LCRS and the behavior of the galaxy numbers counts
as a function of the apparent magnitude \cite{slmp97}.
However we stress again that, due to the lackness of
complete redshift measurements, these evidences are 
statistically weaker than the ones up to one hundred Megaparsec.


\section{Discussion and Conclusions}

Only after clarification of the small scale galaxy correlations
it is possible to investigate the large scale ones.
For what concerns the meaning of the so-called
``correlation lengths'' of galaxies and clusters, 
the behavior of the conditional density
up to $\sim 100 \hmp$ is enough to 
give us the elements for a revision of
both the statistical methods usually used in the 
data analysis as well as the theoretical approach 
(i.e. linear non-linear dynamics, etc.).
We refer the interested reader to \cite{slmp97}
for a more complete discussion of the implications
of the existence of a fractal distribution of matter, at least,
up to $\ 100 \hmp$.

\begin{table} \begin{center}
\caption{\label{tabpre}
The volume limited samples of various 
 catalogs (not still published and analyzed) are characterized by the following 
parameters:
- $R_{VL} (\hmp)$ is the depth of the VL sample considered
with absolute magnitude limit $M_{VL}$
- $\Omega$ is the solid angle
- $R_s (\hmp)$ is the radius of the largest sphere 
that can be contained in the  catalog volume. 
This gives the limit of statistical validity of the sample.
- $r_0(\hmp)$  is the length at which $\xi(r) \equiv 1$.
(distance are expressed in $\hmp$).
}
\begin{tabular}{|c|c|c|c|c|c|}
\hline
     &      &          &    &              &        \\
\rm{Sample} & $\Omega$ ($sr$) & $R_{VL} $ & $M_{VL} $ & $R_s $& $r_0$ \\
     &       &    &    &    &  		         \\
\hline
CfA2  & 1.83  &  101 & -19.5   & 22   & 7    \\
CfA2  & 1.83  &  160 & -20.5   & 36  &  12   \\
SLOAN & $\pi$ &  400 & -19     & 185  &  60  \\
SLOAN & $\pi$ & 600  & -20   &  275 &  90 \\
2dF (South)   & 0.28  &  550 & -19   & 50   &   15  \\
2dF (South)   & 0.28  & 870  & -20   & 100  &  30   \\
      &       &      &       &      &      \\
\hline
\end{tabular}
\end{center} \end{table}

Let us now remark the predictions for future galaxy redshift surveys.
According to the standard interpretation, the length 
$r_0 \simeq 5 \hmp$ characterizes the physical properties of 
galaxy distributions. Therefore deeper samples like CfA2 and SLOAN
should simply reduce the error bar, which is 
now  considered 
to be about $10 \%$. A possible variation of $r_0$ with absolute magnitude,
due to a luminosity bias,  is 
considered plausible but it has 
never been quantified. This should be checked by
varying independently absolute magnitude and depth of the volume limited samples.
However, from this interpretation, the value of $r_0 = 5 \hmp$, 
corresponding to a volume limited of CfA1 with $M=-19.5$, should not change 
when considering in CfA2 and SLOAN volume limited samples with the 
same solid angle $\Omega$ and the same absolute magnitude limit ($M=-19.5$).

In our interpretation, instead,  $r_0$ is  spurious, and it scales 
linearly with the radius $R_s$ of the largest sphere fully contained 
in the volume limited samples. Therefore we predict for the
volume limited sample of CfA2 with $M=-19.5$ (with a solid angle of 
$\Omega=1.1 \; sr$ \cite{par94}) $r_0 \approx 7 \hmp$ (if, in the final 
version of the survey the solid angle is $\Omega =1.8$, 
the value of $R_s$
 increases accordingly, and the value of $r_0$ is shifted up 
to $\sim 9 \hmp$).
Note however that for the deepest 
volume limited CfA2 sample ($M \ltapprox -20$) we 
predict instead $r_{0} \approx 15 \div 20 \hmp$.
For the volume limited sample of the 
full SLOAN with $M=-19.5$
 ($\Omega = \pi$), 
our prediction is
that $r_0 \approx 65 \hmp$. It is clear that 
however, the first SLOAN slice   gives smaller values 
because the solid angle is be small.
In Tab.\ref{tabpre} we report the predictions for $r_0$ in the next future 
surveys.

Finally we would like remark that if it is true that 
the fractal analysis, as good wine, must be taken 
with ``moderation'', it is also true that 
up to now the usual approach in this field 
has suffered of abstinence. We hope that the present discussion
will evolve in a more constructive and cooperative 
way between the field of statistical mechanics
and large scale structure astrophysics, hoping to
drink together some good wine.


\section*{Acknowledgements}
We thank F. Combes, M. Davis, R. Durrer, J.-P. Eckmann,
A. Gabrielli, L. Guzzo, B. Mandelbrot, R. Mali, 
J. Peebles, D. Pfenniger, A. Szalay, and N. Turok, for useful
and interesting discussions, criticisms and comments.
F.S.L. is grateful to N. Sanchez and H. de Vega 
for valuable discussions and for their kind hospitality
at the Observatoire de Paris.

\end{document}